# Thermal conductivity of REBCO tapes with different stabilizers at 4.2 – 200 K

Jun Lu[1*], Yan Xin[1], and Yifei Zhang[2]

[1] National High Magnetic Field Laboratory, Tallahassee, Florida, USA
[2] SuperPower Inc., Glenville, New York, USA

*E-mail: junlu@magnet.fsu.edu

**Abstract.** REBCO coated conductor is a high temperature superconductor that has a wide range of applications, one of which is the current leads of magnet systems. In the design of current leads, it is crucial to minimize their thermal conduction while maintain stable electrical conduction. Therefore, thermal conductivity of various REBCO tapes need to be characterized and analysed. In this research, we measured thermal conductivity of REBCO tapes in the longitudinal direction at 4.2 - 200 K. Samples with Cu, Ag and Ag-3at%Au stabilizers of various thicknesses were measured. The electrical conductivity of these stabilizers was also characterized by residual-resistance-ratio (RRR) measurements and correlated with the thermal conductivity results. We showed that in samples with ≤10 μm Cu stabilizer, thermal conduction is dominated by that of the Cu which has much higher thermal conductivity than the Hastelloy substrate and the superconductor layer. In addition, the sample with 3 μm Ag-3at%Au stabilizer has significantly lower thermal conductivity than that with 3 μm silver stabilizer. It is concluded that REBCO with Ag-3at%Au stabilizer is promising for current lead applications.

## 1. Introduction

REBCO coated conductor is a high temperature superconductor that has wide range of applications in magnets for fusion, high energy physics and high-field research applications. One of the applications is for the high current leads for magnet systems [1]-[3] which allows current injection from the power supply to the superconducting magnet that is at a cryogenic temperature. In the design of current leads, it is desirable to minimize the thermal conduction via REBCO conductor. To reduce the thermal conduction of the current leads, the thickness of the stabilizer layer, typically copper, should be reduced. Further reduction of thermal conduction would require stabilizer made with lower thermal conductivity material. The longitudinal thermal conductivity of REBCO coated conductors with Cu stabilizers have been studied [4] - [9]. This paper focuses on the effect of the stabilizer materials. For this purpose, thermal conductivity of REBCO tapes stabilized by Cu, Ag and Ag-Au alloy is measured. In addition, the residual resistivity ratio of the stabilizers is measured and compared with the values calculated by adding the contributions from ever layer. The small discrepancy between the measured and calculated values will be discussed.

## 2. Experimental

REBCO coated conductor in this work is 4 mm wide, grown on 50 μm Hastelloy by SuperPower Inc. using MOCVD method. After that, a layer of Ag typically 1 - 2 μm per side is deposited. Then the surrounding copper stabilizer of various thickness is electroplated. Figure 1 shows a light micrograph of a cross-section of a REBCO with 40 μm copper (20 μm on each side).

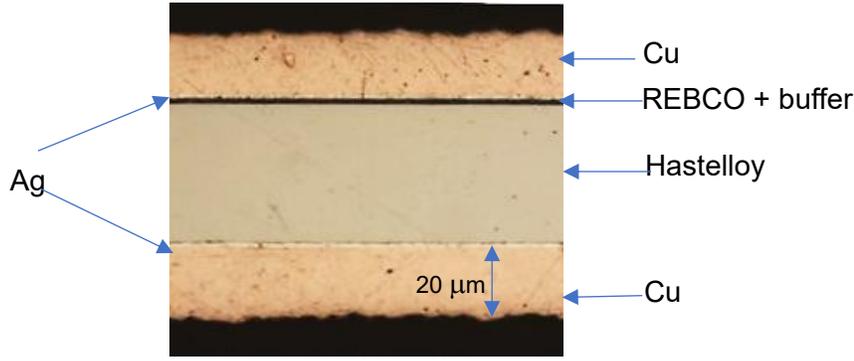

**Figure 1.** Cross-section of a REBCO conductor with 40 μm Cu stabilizer (20 μm per side)

Total of 5 samples were prepared for thermal conductivity measurements. The thickness of the samples is measured by using a digital micrometer. The details of these samples are listed in Table I.

**Table I** REBCO tape samples with different stabilizers

| Sample ID | Stabilizer material | Substrate thickness, μm | Stabilizer thickness, μm | Measured total thickness, μm |
|---|---|---|---|---|
| 1 | Cu | 50 | 100* | 145.8 |
| 2 | Cu | 50 | 40* | 92.8 |
| 3 | Cu | 50 | 10* | 61.2 |
| 4 | Ag | 50 | 3 | 53 |
| 5 | Ag-3at%Au | 50 | 3 | 53.2 |

* The thickness includes 1.5 μm Ag on each side of the tape as shown in Figure 1

The thermal conductivity can be obtained by creating a heat flow through sample and measuring the temperature gradient in the heat flow direction. The thermal conductivity $\kappa$ is calculated by

$$\kappa = W \cdot L / ((A \cdot (T_1 - T_2))） \quad (1)$$

where $W$ is heating power, $A$ is the cross-sectional area, $(T_1-T_2)$ is the temperature difference between the two thermometers which are separated by a length $L$.

In this research, $\kappa$ is measured from 4.2 - 200 K by the thermal transport option of the physical property measurement system (PPMS-TTO) made by Quantum Design Inc. [10]. Each 4 mm wide sample is cut to 18 mm in length. Four thermal leads made of gold-plated copper (cross-section of 0.64 x 0.25 mm²) are attached to the sample by conductive silver-filled epoxy (Epoxy Technology H20E). The epoxy is cured at 120 C for 15 minutes, which should not cause significant change in transport properties of the stabilizers by heating [11]. Two Cernox thermometers and a heater are attached to the sample's thermal leads by pressure contacts of set screws. A picture of the sample mounted on the measurement puck is shown in Figure 2. The typical length between the two inner leads is 8 mm. During a thermal conductivity measurement, the sample is cooled down to 3 K. Then the measurement is taken in continuous mode. That is while the system temperature (the cold end temperature) is slowly ramping up, the heater is fired for a period long enough that the response of the two thermometers (in transient) can be simulated to get the steady-state $(T_1-T_2)$. Then the thermal conductivity is obtained by (1). The measurement repeats until reaches the end of the temperature ramp. The system ramp rates are 0.2 K/min from 3 – 20 K, and 0.5 K/min from 20 – 300 K.

The measurement is taken place in high vacuum (pressure < 1 x 10$^{-6}$ Torr). This is to ensure that the heat loss due to gas conduction is negligibly small. The radiation heat loss is calculated by,

$$P_{rad} = \sigma_T \times (S/2) \times \varepsilon \times (T_1^4 - T_2^4) \qquad (2)$$

Here $\sigma_T$ is the Steffen-Boltzmann constant which is 5.67 × 10$^{-8}$ W m$^{-2}$ K$^{-4}$, $S$ is the sample's total surface area, $\varepsilon$ is the infrared emissivity of the material which is estimated to be 0.3 for unpolished metal surface with significant roughness. Although the radiation loss is corrected, the uncertainty in emissivity is significant at temperatures above 200 K.

The measurement error comes from the following three main sources, the error in distance of two thermal leads, the fitting error of the $\Delta T$ vs time, and the radiation loss. The total measurement error of thermal conductivity is estimated to be less than 10% for all the temperatures.

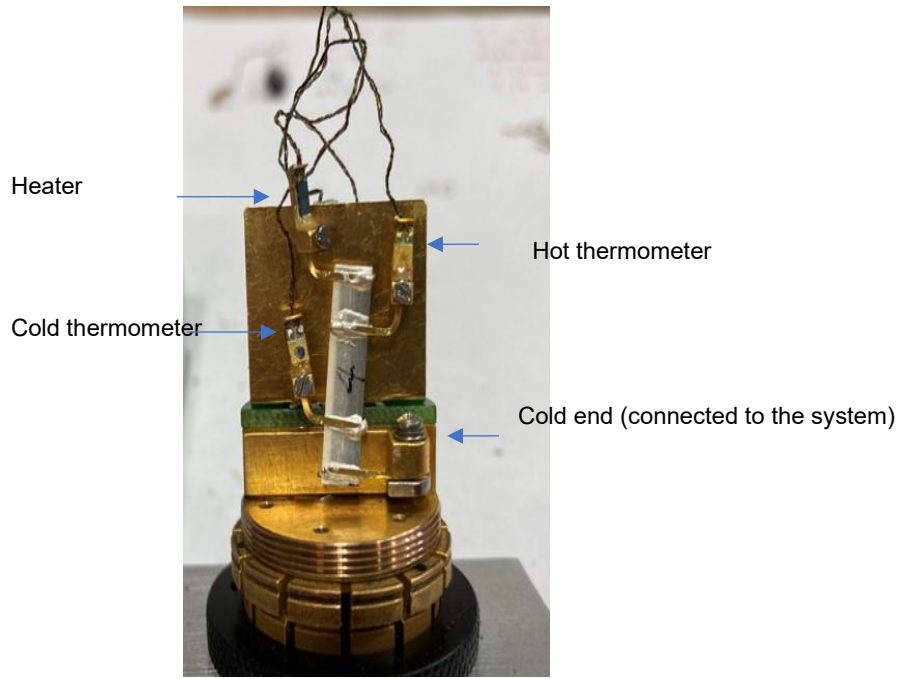

**Figure 2.** A sample is mounted on the PPMS-TTO puck.

The residual resistance ratio (RRR) is defined as the ratio of resistance at 295 K and that at 4.2 K. It is measured for the stabilizers of all 5 samples. For this measurement, the Cu stabilizers of sample 1 - 3 are peeled from the substrate, and etched by a diluted nitric acid solution of $HNO_3$: $H_2O$ = 3 : 500 to remove the residual superconductor. For sample 4 and 5, since the stabilizer is too thin to peel, it is removed from the REBCO side by chemical etching with $HNO_3$: $H_2O$ = 1 : 1, while the stabilizer on the substrate side is protected by a Kapton adhesive tape. This is followed by removal of

REBCO layer by using the above mentioned nitric acid solution. The stabilizer on the substrate side is preserved and its RRR measured together with the substrate. RRR of the stabilizer can be calculated by thicknesses of the substrate and stabilizer, the resistivity of the substrate at 295 K and 4.2 K, and resistivity of the stabilizer at 295 K.

The resistance of the stabilizers is measured by a four-probe method using a HP 6631B as current source and a Keithley 2010 as digital voltmeter. The measurement is performed at room temperature (295 K) and in liquid helium bath (4.2 K) respectively. The ratio of the two is defined as RRR.

For stabilizer characterization including cross-section cutting by focused-ion-beam and chemical analysis by energy dispersive spectroscopy (EDS), a Thermal Scientific Helios G4 UC dual-beam field-emission SEM is used. The cross-section of the sample was by coating a protective layer of platinum then cut a trench by the focused ion beam to expose the cross-section. The cross-sectional image was taken with electron beam incident angle of 52 degrees.

### 3. Results and discussions

*3.1 The Ag and Ag-Au stabilizer characterization*

For the thin stabilizer layers of Ag and Ag-Au alloy, scanning electron microscopy was used to characterize them. A typical image is shown in Figure 3. The total thickness Ag and Ag-Au stabilizers are about 3 μm (1.5 μm per side). The composition of Ag-Au stabilizer is characterized by EDS to be Ag-3at%Au.

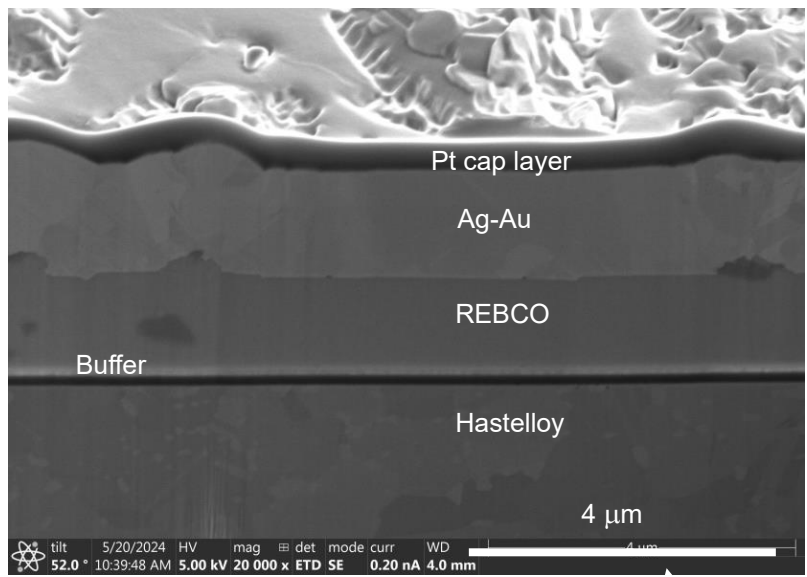

**Figure 3.** SEM examination of the Ag-Au coating on sample 5.

*3.2 RRR of stabilizers*

The measured RRR of the stabilizers of each sample is listed in table II. For Cu stabilizers, RRR increases with its thickness. This is consistent with [11] where RRR increases with Cu grain size which increases with thickness of electroplated Cu film. The RRR of the thin Ag and Ag-3at%Au is comparable with [11] and [13] respectively.

**Table II** RRR of the stabilizers

| Sample ID | Stabilizer material | Stabilizer thickness, μm | Measured stabilizer RRR | Simulated stabilizer RRR |
|---|---|---|---|---|
| 1 | Cu | 100 | 133 | 108 |
| 2 | Cu | 40 | 58 | 63 |
| 3 | Cu | 10 | 31 | 25 |
| 4 | Ag | 3 | 12.8 | 8 (Cu) |
| 5 | Ag-3at%Au | 3 | 2.3 | - |

*3.3 Thermal conductivity*

The measured thermal conductivity is presented in Figure 4. For samples with ≥10 μm Cu stabilizers typical for magnet applications, thermal conductivity increases with the Cu stabilizer thickness. This is expected because with the same substrate thickness, the higher fraction of the Cu in REBCO leads to higher thermal conductivity, since Cu is much better thermal conductor than the substrate. For samples with 3 μm stabilizers typically used for current lead applications, Ag-3at%Au stabilized REBCO has significantly lower thermal conductivity than Ag stabilized one.

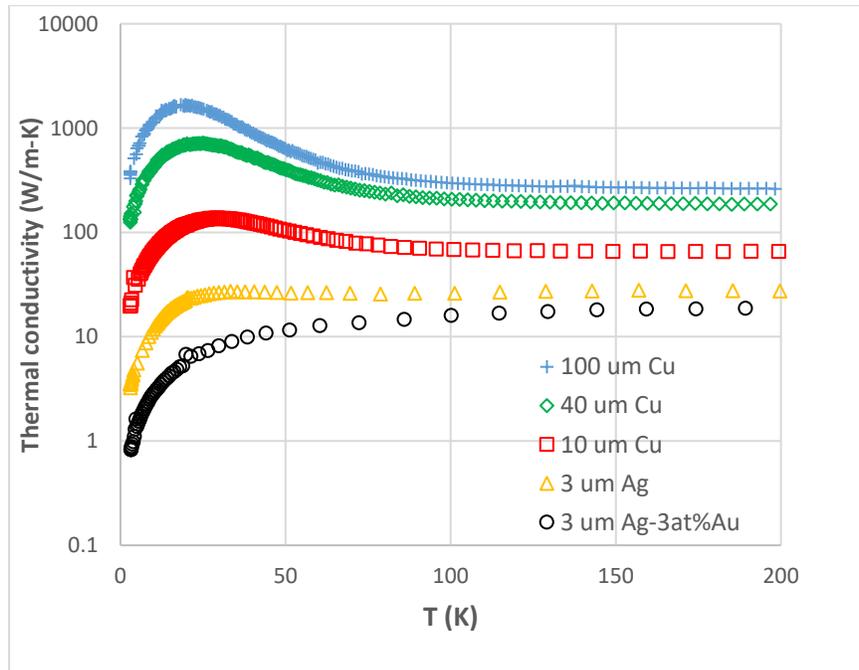

**Figure 4.** Measured thermal conductivity of samples with different stabilizers.

To explain the observed thermal conductivity of each sample, we consider the thermal conduction of each layer in the coated conductor. In a layered composite like REBCO, the longitudinal heat conduction of each layer is in parallel. The total thermal conductance $K$ is the sum of that of each layer,

$$K = \kappa \cdot A/L = \kappa_1 \cdot A_1/L + \kappa_2 \cdot A_2/L + \ldots \quad (3)$$

where $L$ is the same length, $\kappa$ and $A$ are the thermal conductivity and cross-sectional of the composite. $\kappa_i$ and $A_i$ are the thermal conductivity and cross-sectional area of each layer. Therefore,

$$\kappa = \Sigma \kappa_i (A_i/A) \quad (4)$$

We denote $t$ and $t_i$ as the thickness of the composite and the $i^{th}$ layer respectively. Since all layers have the same width, cross-sectional area ratio $A_i/A = t_i/t$,

$$\kappa = \Sigma \kappa_i (t_i/t) \quad (5)$$

From Equation (5), for a layer $i$ with low thermal conductivity and small thickness, $k_i t_i$ is small. Its contribution to the total thermal conductivity will be negligibly small. For this reason, in the following discussions we ignore the contribution by the buffer layer because of its small thickness about 0.3 µm (for example, as shown in Figure 2), which is less than 0.6% of the total thickness even for the samples with only a 3 µm stabilizer.

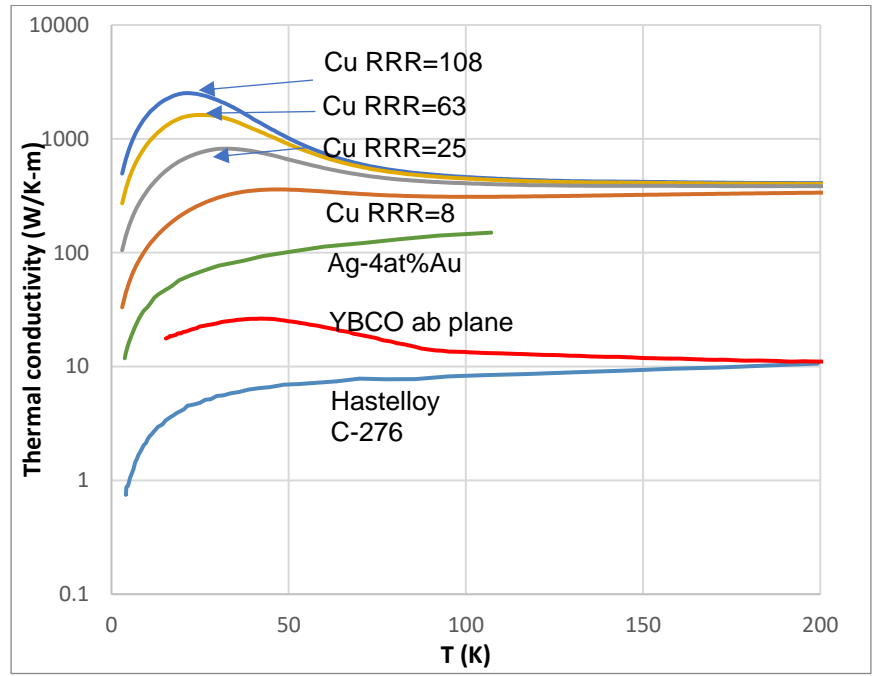

**Figure 5.** Material thermal conductivity data. Thermal conductivity of Cu is calculated by an empirical formula in [12]; that of Ag-4at%Au is reproduced from [13]; that of Hastelloy C-276 is reproduced from [14]; and that of YBCO in its ab plane is reproduced from [15].

We now compare the measured data with calculated ones using equation (5). In the calculation thermal conductivity and thickness for each layer are needed. The reference thermal conductivities of Cu [12], Ag-4at%Au [13], Hastelloy [14] and YBCO [15] are reproduced and presented in Figure 5. Since the thermal conductivity of Cu strongly depends on its RRR, RRR is used to calculate thermal conductivity of Cu using an empirical formula given by Ref. [12]. Its value is adjusted to simulate the measured thermal conductivity vs. $T$ curve. The simulated RRR are listed in Table II.

The comparison between experimental $\kappa$ (T) curve and the simulated ones are shown in Figure. 6. In cases of 100 or 40 μm copper stabilizer (Figure 6(a)), the agreement is very good. The contribution of Hastelloy is negligibly small, therefore is not plotted. For the sample with 10 μm Cu shown in Figure 6(b), the agreement is satisfactory with noticeable contribution from Hastelloy substrate. For REBCO with 3 μm Ag stabilizer (Figure 6(c)), the contribution from the Hastelloy substrate is significantly more pronounced. The simulation is difficult because thermal conductivity data for Ag of low RRR are not available. We attempted to use Cu $\kappa$ (T) data (RRR = 8) for the simulation based on the similarity of thermal conductivity behavior between Cu and Ag. As shown in Figure 6(c), the general feature of simulated curve is similar with the measured data. The significant discrepancy especially at 30 – 50 K can be attributed to the difference between Ag and Cu. For Ag-3at%Au stabilizer, available data for Ag-4at%Au from 4 – 100 K are used in the simulation. It should be noted that even in this case of low total thermal conductivity, the contribution from the superconductor layer is still very low as shown in Figure 6(d). Given the appreciable uncertainty in the calculated thermal conductivity of Ag-3at%Au, the agreement between the measured and the calculated values is reasonably good.

It is noticed that the RRR obtained from thermal conductivity simulation is notably different from the electrically measured value both shown in Table II. Similar discrepancies were observed by Bonura and Senatore [5] where RRR determined by resistivity measurement was somewhat different from those determined by thermal conductivity measurement which was attributed to the uncertainty in Cu thermal conductivity calculation by Ref. [12]. In addition, the nonuniformity in stabilizer thickness across the sample width may contribute to the error in the simulated thermal conductivity.

Our data show that Ag-Au alloy stabilizer significantly reduces the thermal conductivity of REBCO tapes. However, Ag layer is necessary for REBCO oxygenation annealing. Adding 3 at% Au to Ag could have impact on the REBCO performance especially at the metal-superconductor interface. Ekin et al [16] studied the interfacial resistance of Ag-YBCO and Au-YBCO. Both can achieve very low interfacial resistance but with different annealing temperatures. So, it is conceivable that the annealing temperature needs to be optimized for Ag-3at%Au to achieve the lowest interfacial resistance.

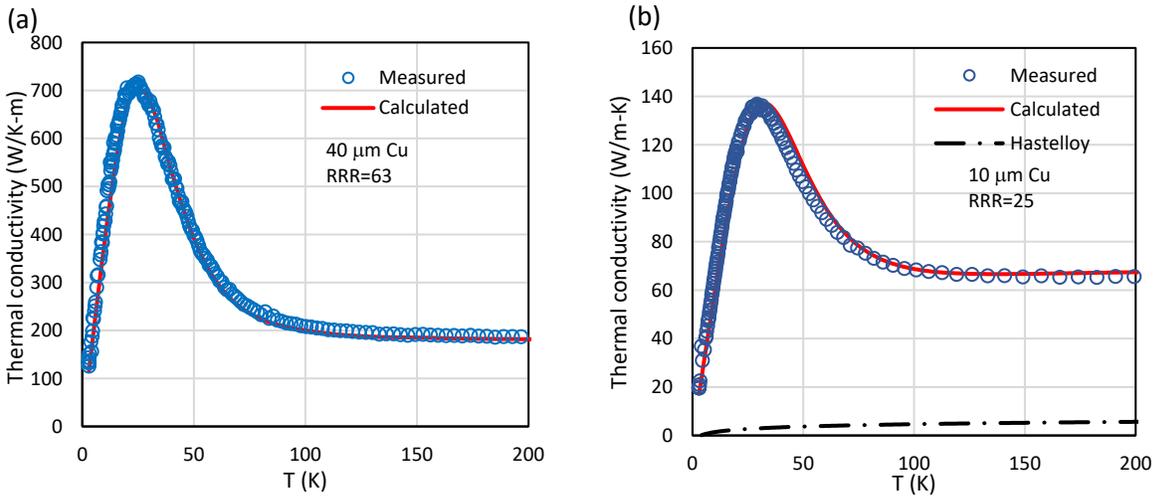

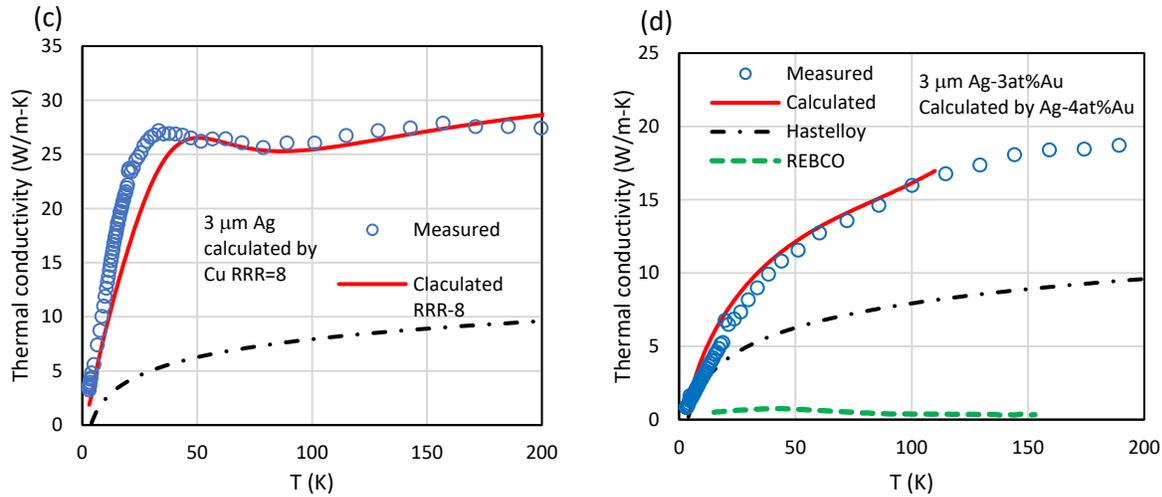

**Figure 6.** Comparison of measured and simulated thermal conductivity. (a) the sample with 40 μm Cu of RRR = 63. The contributions from Hastelloy substrate are negligibly small, therefore, not shown. (b) REBCO with 10 μm Cu of RRR = 25, the contribution of Hastelloy cannot be disregarded. (c) REBCO with 3 μm Ag stabilizer. The simulation uses data for Cu of RRR = 8, (d) REBCO with 3 μm Ag-3at%Au. The simulated curve uses data for Ag-4at%Au in Ref. [9].

## 4. Conclusion

We measured thermal conductivity of REBCO tapes with different stabilizers at temperatures between 4.2 K and 200 K using the thermal transport option of a physical property measurement system (TTO-PPMS). The electrical conductivity of the stabilizers was also characterized by residual-resistance-ratio (RRR) measurements. The measured RRR values are comparable with those obtained from thermal conductivity simulation. Our results show that the thermal conductivity of REBCO tapes with a Cu stabilizer thicker than 10 μm is dominated by that of the Cu layer. In such cases, the contribution of the 50 μm Hastelloy substrate is negligible. For samples with 3 μm stabilizers intended for current leads applications, Ag-3at%Au stabilizer has significantly lower thermal conductivity than that with a Ag stabilizer. It is concluded that Ag-3at%Au stabilizer is promising for current leads applications where minimal thermal conduction is desirable.

## 5. Acknowledgement

This work was performed at the National High Magnetic Field Laboratory, which is supported by National Science Foundation Cooperative Agreement No. DMR-1644779, and the State of Florida.